\documentstyle[11pt,epsfig]{article}
\newcommand{\be}{\begin{equation}}
\newcommand{\ee}{\end{equation}}
\newcommand{\beqn}{\begin{eqnarray}}
\newcommand{\eeqn}{\end{eqnarray}}

\newcommand{\spc}[1]{\mbox{\hspace{#1}}}
\newcommand{\frho}{\mbox{\boldmath $\rho$}}
\newcommand{\qf}{{\bf q}}
\newcommand{\kf}{{\bf k}}

\newcounter{savefig}

\begin{document}
\begin{flushright} hep-ph/9509303 \end{flushright}
\begin{center}
\begin{Large}
{\bf Conformal Invariance of the Transition Vertex \\
$2 \rightarrow 4$  gluons}\\
\end{Large}
\vspace{0.5cm}
J. Bartels$^a$,
L.N.Lipatov$^{b\; c}$ \footnote{Alexander von Humboldt Preistr\"ager},
                       M.W\"usthoff$^a$ \\
\end{center}
\vspace{0.5cm}
$^a${\it II. Institut f\" ur Theoretische Physik,
Universit\" at Hamburg.}
\\
$^b${\it Deutsches Elektronen Synchrotron, DESY, Hamburg.}
\\
$^c${\it St.Petersburg Nuclear Physics Institute, 188350, Gatchina,
Russia}
\vspace{2.0cm}

\noindent
{\bf Abstract:} We show that the transition vertex: two reggeized gluons
$\to$ four reggeized gluons is invariant under M\"{o}bius
transformations. This provides an important step in defining a
conformally invariant effective field theory for QCD in the Regge limit.
\vspace{2cm}

\section{Introduction}
\setcounter{equation}{0}
\noindent
The Balitsky-Fadin-Kuraev-Lipatov (BFKL) ~\cite
{BFKL} Pomeron has recently
attracted much interest, in particular in connection with the small-x
behavior of the deep inelastic structure function of the proton.
The BFKL Pomeron represents, within QCD, the leading-logarithmic
approximation of the Regge limit (energy $s/Q^2 \approx 1/x_B
\rightarrow \infty$, momentum transfer
$t$ small and fixed). Since at large energies it violates the
unitarity bound, there is no doubt that subleading corrections to the
BFKL
Pomeron are of vital importance, and strategies have to be found
which allow the summation of all these contributions. \\ \\
In ~\cite{L86,L93} it has been observed that the kernel of the BFKL
Pomeron is invariant under M\"obius transformations: in particular,
starting from the Fourier transform of the BFKL equation, it has been
shown that the equation remains invariant under the inversion
$\rho \to 1/ \rho$ ($\rho$ denotes the complex variable $\rho_x + i
\rho_y$ where $\rho_x$, $\rho_y$ are the two components of the impact
parameter). Moreover, when writing the BFKL evolution equation
in the form of the Schroedinger equation, one can show
that the Hamiltonian has the property of holomorphic separability, i.e.
it can be written as a sum of two pieces which depend only upon $\rho$
and $\rho^*$, resp. Both these properties can be derived most
elegantly from the following representation of the Hamiltonian \cite{L93}:
\beqn
\cal H &=&
           \frac{g^2 N_c}{8\pi^2} \left(
           \frac{1}{p_1 p_2^* } \,\, \ln | \rho_{12}|^2 \,\,
     p_1 p_2^* + \mbox{h.c.}
+\ln |p_1|^2 + \ln |p_2|^2 - 4 \psi(1) \right),
\eeqn
where $ p_r = i \frac{\partial}{\partial \rho_r}$.
This Hamitonian can be split into two pieces:
\beqn
{\cal H} = H+H^*
\eeqn
with
\beqn
 H=  \frac{g^2 N_c}{8\pi^2} \left(
      \frac{1}{p_1} \,\, \ln \rho_{12} \,\, p_1
    + \frac{1}{p_2} \,\, \ln \rho_{12} \,\, p_2
     + \ln(p_1 p_2) - 2 \psi(1) \right).
\eeqn
After some algebra one finds that (1.3) is identical with:
\beqn
H=         \frac{g^2 N_c}{8\pi^2} \left(
  \ln(\rho_{12}^2 p_1) + \ln (\rho_{12}^2 p_2)
     - 2 \ln \rho_{12} - 2 \psi(1) \right).
\eeqn
(both representations (1.3) and (1.4) are, strictly speaking,
somewhat symbolic as long as one does not specify
the boundary conditions for the integral operators).

Eq.(1.2) means that the Hamiltonian has the property of
holomorphic separability, and the invariance under
M\"obius transformations (in particular the inversion $\rho \to 1/\rho$)
can be deduced from (1.4). The importance of these properties is
rather obvious: the separability reduces the two-dimensional problem to
a one-dimensional one (instead of quantum mechanics in a plane one is
dealing with quantum mechanics on a line), and the symmetry
allows to solve the eigenvalue problem in terms of the
representations of the M\"obius group. In this way it was possible to
solve the BFKL equation also in the nonforward direction.
\\ \\
As far as the
restoration of unitarity is concerned it is believed that the most
important corrections will come from diagrams with a large number
of (reggeized) gluons in the t-channel. As a first step, one considers
diagrams ~\cite{BKP} where $n$, the number of gluons in the t-channel, is
conserved. The high energy behavior of these contributions is given
by the energy spectrum of a quantum mechanical system
of n particles with pairwise interactions through the BFKL kernel.
As a result of the conformal symmetry of the BFKL
Hamiltonian (1.1) - (1.4), it is tempting to expect that the problem of
determining the energy spectrum may be exactly soluble: in fact,
for the large-$N_c$ limit it has been shown ~\cite{L94,FK} that the
system is
equivalent to the XXX Heisenberg model for spin zero, and the energy
behaviour can, at least in principle, be computed analytically.\\ \\
As the next step in this direction, one has to adress the question
whether there are kernels which change the number of t-channel gluons.
For the simplest case, the transition from 2 to 4 gluons, a new
kernel has been derived in ~\cite{B,BW}, which, similar to the BFKL
kernel, consists of several pieces. The fact that such a number-changing
vertex exists means that the quantum mechanical problem (``no particle
production``) studied in ~\cite{L94,FK}, in fact, will turn into a
quantum field theory in 2+1 dimensions. In view of the conformal
symmetry which, in the case of the n-gluon system, has turned out to
be so powerful, we are then facing the question whether the same
symmetry is preserved also by the new transition vertex. If so, there is
hope that even the field theory may have an analytic solution.
It is the purpose of this paper to show that, in fact, the new
transition vertex is invariant under conformal (M\"obius)
transformations. \\ \\
To complete this short overview of strategies beyond the BFKL
approximation, it is important to note that both the BFKL kernel (which
contains, as a part of it, also the gluon trajectory function) and
the transition vertex have so far been computed only in leading
order of $\alpha_s$. For the BFKL kernel, calculations of the
$\alpha_s^2$ corrections are partly been done ~\cite{FL}. Interest in
these corrections also comes from the recent observation ~\cite{EKL,EHW,
BF} that a resummation of the singular terms in the quark and gluon
anomalous dimension seems to play a very crucial role in the GLAP
evolution at small x. A field theoretic formulation (effective action)
which takes into account these nonleading corrections to all orders
has recently been suggested in ~\cite{L95,L91}.\\ \\
This paper is organized as follows. Since the transition vertex cannot
be brought into a compact form analogous to (1.1) which would allow to
derive both symmetry and separability in a rather elegant and
straightforward fashion, we first (section 2) review the explicit proof
of conformal invariance of the BFKL kernel. We then (section 3) define
the transition vertex for $2 \to 4$ gluons and present a brief scetch of
its derivation. The proof of the conformal invariance is contained in
section 4, and in section 5 we conclude with a brief summary.\\ \\
\section{Conformal Invariance of the BFKL Kernel}
\setcounter{equation}{0}
\subsection{Fourier Transform of the BFKL Kernel}
In this section we briefly repeat, as an illustration and as
a preparatory step, the proof of conformal invariance of the BFKL
kernel (Fig.1). The (non-amputated) amplitude will be denoted by
$\Phi_{2} (q_1,q_2)$ (we suppress the dependence upon angular momentum
$\omega$). We first discuss the Fourier transform and then prove the
invariance under the inversion $\rho \rightarrow 1/\rho$.\\ \\
It will be convenient to introduce complex coordinates in the
two-dimensional transverse space:
\beqn
\qf &=& (q_x,q_y) \\
q=q_x+i q_y \;\; &,& \;\; q^{\ast}=q_x-iq_y \\
\qf\cdot\frho &=& \frac{1}{2}( q \rho^{\ast}+q^{\ast}\rho).
\eeqn
The gluon production vertex then reads (in the light-cone gauge of
{}~\cite{L91})
\be
{\cal C}_1 = -2 q_1 q_{1'}^{\ast} \frac{1}{(q_{1'}-q_1)^{\ast}},
\ee
and for the squared production amplitude one obtains
\be
{\cal C}_{\mu}{\cal C}^{\mu} =
 \frac{q_1 q_{1'}^{\ast}q_{2'}q_2^{\ast}}
{(\qf_{1'}-\qf_1)^2}  + \mbox{h.c.} .
\ee
After multiplication with the inverse propagators for the upper gluons
the integral in Fig.1 takes the well-known form:
\beqn
|q_{1}|^2|q_{2}|^2 \left( {\cal K}
 \Phi_{2} \right)(\qf_1,\qf_2) =
\int d^2 \qf_{1'}
\left(
\frac{q_1 q_{1'}^{\ast}q_{2'}q_2^{\ast}}
{|q_{1'}-q_1|^2+\lambda^2}  + \mbox{h.c.} \right)
\Phi_{2}(\qf_{1'},\qf_{2'})  \nonumber \\
- \pi \left( \ln \frac{|q_1|^2}{\lambda^2}
+ \ln \frac{|q_2|^2}{\lambda^2} \right)
|q_{1}|^2|q_{2}|^2 \Phi_{2}(\qf_{1},\qf_{2}).\hspace{1cm}
\label{bfkl2}
\eeqn
Here we have introduced the infrared regulator $\lambda^2$ (gluon mass),
which later on we shall remove again. Assuming that $\Phi_2$ is
invariant under M\"obius transformations, we shall prove
that the integral on the rhs remains invariant. \\ \\
The Fourier transformation is defined as follows
\beqn
\Phi_{2}(\frho_1,\frho_2) =
\int \frac{d^2 \qf_1 d^2 \qf_2}{(2\pi)^4}
\exp \left[ i\qf_1 \frho_1 + i \qf_2 \frho_2 \right]
\Phi_{2}(\qf_{1},\qf_{2}),
\eeqn
and we begin with the gluon trajectory function. Inserting
\be
1= \frac{1}{(2 \pi)^2} \int d^2 \frho_0  \int d^2 \kf
\;e^{i\frho_0(\kf-\qf_1)}
\ee
we obtain
\beqn
- \Delta_1 \Delta_2
 \int d^2 \frho_0    \left[ \frac{1}{(2 \pi)^2}\int d^2 \kf
e^{i\kf(\frho_1-\frho_0)} \pi \ln \frac{|k|^2}
{\lambda^2} \right]
\Phi_{2}(\frho_{0},\frho_{2}).
\eeqn
For the integral in brackets we introduce the ultraviolet cutoff
parameter $\epsilon$ which will be removed at the end of our
calculations:
\beqn
- \left[ \frac{1}{(2 \pi)^2}\int d^2 \kf
e^{i\kf(\frho_1-\frho_0)} \pi \ln \frac{|k|^2}
{\lambda^2} \right] = \Gamma(\frho_1-\frho_0)  \\
\Gamma(\frho_1-\frho_0) = \frac{1}{|\frho_{10}|^2}
\theta(|\frho_{10}|-\epsilon) +c(\epsilon,\lambda)
\delta^{(2)}(\frho_{10}).
\eeqn
The function $c(\epsilon, \lambda)$ is obtained from the integral
\beqn
\int d^2 \frho \frac{1}{|\rho|^2} \theta ({|\rho|-\epsilon})
e^{-i\kf\frho}&=& \int_0^{2\pi}d\varphi \int_{\epsilon}^{\infty}\frac{ d|\rho|}
{|\rho|} \exp{(-i|k||\rho|\cos \varphi)} \nonumber \\
&=& 2 \pi \left[\psi(1) - \ln (\epsilon |k|)\right]
- 4 \int_0^{\pi/2}d\varphi\ln(\cos \varphi) \;+\;O(\epsilon)
\nonumber \\& =&
2 \pi \left[\psi(1) + \ln \frac{2}{\epsilon |k|} \right]\;+\;
     O(\epsilon),
\eeqn
where the series representation of the Exponential-integral has been used
neglecting the terms proportional to $\epsilon$.
We then find for $c(\epsilon,\lambda)$:
\be
c(\epsilon,\lambda)= 2\pi [  \ln \lambda + \ln
\frac{\epsilon}{2} -  \psi(1)].
\label{traj}
\ee
\\
Next we turn to the Fourier transform of the first term on the rhs of
eq.(2.6), the square of the production vertex. We make use of the
complex coordinates, defined in eqs.(2.1) - (2.3). With
$\partial = \partial / \partial \rho$, $\partial^* = \partial /
\partial \rho^*$, $\Delta = 4 \partial \partial^*$ we find:
\beqn
&&\int \frac{d^2 \qf_1 d^2 \qf_2}{(2 \pi)^4}
e^{i\qf_1\frho_1 + i\qf_2\frho_2}
\int d^2 \qf_{1'}
\left(
\frac{q_1 q_{1'}^{\ast}q_{2'}q_2^{\ast}}
{|q_{1'}-q_1|^2+\lambda^2}  + \mbox{h.c.} \right)
 \Phi_{2}(\qf_{1'},\qf_{2'}) \vspace{5cm}
\nonumber \\ &=&
\int d^2 \frho_{1'} d^2 \frho_{2'} \int \frac{d^2 \qf_1 d^2 \qf_2}{(2\pi)^4}
d^2 \kf \;\frac{16}
{|k|^2+\lambda^2}\;
\partial_1^{\ast} \partial_2\;
e^{-i\kf(\frho_{1'}-\frho_{2'})} \;\;\cdot\nonumber\\
&&\spc{3cm}\cdot\;\;e ^{i\qf_1(\frho_1-\frho_{1'})+i\qf_2(\frho_2-\frho_{2'})}
\;\partial_{1'}\partial_{2'}^{\ast}
\Phi_{2}(\frho_{1'},\frho_{2'})\;+\;\mbox{h.c.} \\
&=& 16 \;
\left(\partial_1^{\ast} \partial_2
\int d^2 \kf
\frac{1}{|k|^2+\lambda^2}
e^{-i\kf(\frho_{1}-\frho_{2})}
\partial_{1} \partial_{2}^{\ast}\;+\; \mbox{h.c.}\right)
\Phi_{2}(\frho_{1},\frho_{2}).
\hspace{2.1cm}\nonumber
\eeqn
The $\kf$ - integral gives
\be
2 \pi [ \ln \frac{2}{|\frho_{12}|}  -
\ln \lambda + \psi(1)],
\label{f1}
\ee
where we have used the integral:
\beqn
\int d^2 \kf \frac{e^{i\kf\frho}}{\kf^2+\lambda^2} &=&
\int_0^{2\pi}d\varphi \int_0^{\infty} d |k| \, \frac{|k|
\exp{(i|k||\rho|\cos \varphi)}}{|k|^2+\lambda^2} \hspace{2cm} \nonumber \\
&=&\frac{1}{2} \int_0^{2\pi}d\varphi \int_0^{\infty} d |k|
\exp{(i|k||\rho|\cos \varphi)}\left(\frac{1}{|k|+i\lambda} +
\frac{1}{|k|-i\lambda}\right) \\& =&
2 \pi \left[\psi(1) +  \ln \frac{2}{|\rho|\lambda}\right] +O(\lambda)\;\;.
\nonumber\eeqn
Moving from the second to the third line in eq.(2.16) we have shifted
the integration variable $|k|$ by $- i\lambda$ in the frist and $+i\lambda$
in the second term. After that the series
representation of the Exponential-integral similarly to eq.(2.12) was used.
Comparing with eq.(2.9)-(2.13) we find that
$\;\psi(1)-\ln \lambda +\ln 2\;$ cancels between the
trajectory and the production vertex, and our eq.(2.6)
in configuration space reads \footnote{We have
taken out a factor 16 on both sides}:
\beqn
&&|\partial_1|^2|\partial_2|^2
({\cal K}\Phi)(\frho_1,\frho_2) =
|\partial_1|^2|\partial_2|^2 \int \frac{d^2 \frho_0}
{|\rho_{10}|^2} \theta(|\rho_{10}| - \epsilon)
\Phi_{2}(\frho_{0},\frho_{2})\nonumber \\
&&\spc{1cm}+\; |\partial_1|^2|\partial_2|^2 \int \frac{d^2 \frho_0}
{|\rho_{20}|^2} \theta(|\rho_{20}| - \epsilon)
\Phi_{2}(\frho_{1},\frho_{0})  \\
&&\spc{1cm}+\; 4 \pi \ln \epsilon
|\partial_1|^2|\partial_2|^2 \Phi_{2}
(\frho_{1},\frho_{2}) \;
-\; \left[2\pi \partial_1^{\ast}\partial_2 \ln |\rho_{12}|
\partial_1 \partial_2^{\ast}+ \mbox{h.c.} \right]
\Phi_{2}(\frho_{1},\frho_{2}). \nonumber
\eeqn
We still rearrange the differential operators in the first two terms,
using partial integration
\beqn
\partial_1 \partial_1^{\ast} \int d^2 \frho_0 g(|\frho_{10}|^2)
\Phi_{2}(\frho_{0},\frho_{2})
= \partial_1 \int d^2 \frho_0 \partial_1^{\ast}
g(\rho_{10}\rho_{10}^{\ast})
\Phi_{2}(\frho_{0},\frho_{2})
\hspace{1.2cm} \nonumber \\
=\partial_1 \int d^2 \frho_0 (-1) \partial_0^{\ast}
g(\rho_{10}\rho_{10}^{\ast})
\Phi_{2}(\frho_{0},\frho_{2})
=\partial_1 \int d^2 \frho_0 g(|\frho_{10}|^2) \partial_0^{\ast}
\Phi_{2}(\frho_{0},\frho_{2}).
\eeqn
Distributing the derivatives in a symmetric way we end up with
\beqn
&&2 \; |\partial_1|^2|\partial_2|^2
({\cal K}\Phi)(\frho_1,\frho_2) \;=\;
\partial_1 \partial_2^{\ast}
\int \frac{d^2 \frho_0}{|\rho_{10}|^2} \theta(|\rho_{10}| - \epsilon)
\partial_0^{\ast} \partial_2
\Phi_{2}(\frho_{0},\frho_{2}) \nonumber \\
&-&2 \pi \partial_1 \partial_2^{\ast} \ln|\rho_{12}|
\partial_1^{\ast} \partial_2\;+\; 2\pi \ln \epsilon
|\partial_1|^2|\partial_2|^2 \Phi_{2}(\frho_1,\frho_2)
+\; \mbox{h.c.}\; +\; \left[ 1 \leftrightarrow 2 \right].
\label{abs}
\eeqn
Due to the logarithmic divergence at $\frho_0 = \frho_1$
any change of the argument of the $\theta$-function
has to be compensated by an appropriate logarithmic term.
In the case of eq.(2.19) the following equality applies:
\be
 \int \frac{d^2 \frho_0}{|\rho_{10}|^2}
\theta(|\rho_{10}|-\epsilon)  f({\frho_0})
\;=\;\int \frac{d^2 \frho_0}{|\rho_{10}|^2}
   \theta(\frac{|\rho_{10}|}
 {|\rho_{12}|}-\epsilon) f({\frho_0})
      +\;2\pi \ln |\rho_{12}| f({\frho_1}),
\ee
where $f(\frho_0)$ denotes the remainder of the integrand in (2.19)
which is regular near $\frho_0 = \frho_1$.
The kernel finally achieves the form
\footnote{This form has first been used in ~\cite{GLN} for the Odderon
 problem in QCD.}:
\beqn
&& 2 \; |\partial_1|^2|\partial_2|^2
({\cal K}\Phi)(\frho_1,\frho_2) \;=\;
\partial_1 \partial_2^{\ast}
\int \frac{d^2 \frho_0}
{|\rho_{10}|^2} \;\theta(\frac{|\rho_{10}|}
{|\rho_{12}|} - \epsilon)\;
\partial_0^{\ast} \partial_2
\Phi_{2}(\frho_{0},\frho_{2}) \\
&&\spc{1cm}+ \;2\pi \ln \epsilon
|\partial_1|^2|\partial_2|^2 \Phi_{2}(\frho_1,\frho_2)
\;+\; \mbox{h.c.} \;+\; \left[ 1 \leftrightarrow 2 \right].
\nonumber
\eeqn
\subsection{Invariance under Inversion}
Now we want to verify that the configuration space kernel
is invariant under conformal transformations. It is rather obvious
that the kernel in (2.21) is invariant under dilatations,
translations and rotations. The only property to be shown
is the invariance under inversions.
Let us perform the transformation
\beqn
\rho_i & \rightarrow & \frac{1}{\rho_i}
\label{trans} \\
\partial_i & \rightarrow & -\rho_{i}^2 \partial_{i}
\\
d^2 \rho_0 & \rightarrow & \frac{d^2 \rho_{0}}
{|\rho_{0}|^4} \\
|\rho_{i0}|^2 & \rightarrow &  \frac{|\rho_{i0}|^2}
{|\rho_{i}|^2|\rho_{0}|^2}
\eeqn
and apply this transformation to
the integral kernel. We obtain
\beqn
&&2\;  |\rho_{1}|^4 |\rho_{2}|^4
 |\partial_{1}|^2|\partial_{2}|^2
({\cal K}\Phi)(\frho_{1},\frho_{2})\nonumber\\
&&\spc{1cm}=\;\rho_{1}^2 \rho_{2}^{\ast \; 2}
\partial_{1} \partial_{2}^{\ast}
\int \frac{d^2 \frho_{0}}
{|\rho_{10}|^2}
\frac{|\rho_{1}|^2|\rho_{0}|^2}{|\rho_{0}|^4}
 \theta(\frac{|\rho_{10}||\rho_{2}| }
{|\rho_{12}||\rho_{0}| } - \epsilon)
\rho_{0}^{\ast \; 2} \rho_{2}^2
\partial_{0}^{\ast} \partial_{2}
\Phi_{2}(\frho_{0},\frho_{2}) \nonumber \\
&&\spc{2cm}+\; 2\pi \ln \epsilon
 |\partial_{1}|^2|\partial_{2}|^2
 \Phi_{2}(\frho_{1},\frho_{2})
\;+\;\mbox{h.c.}\;+\; \left[ 1 \leftrightarrow 2 \right]
\nonumber \\
&&\spc{1cm}=\;
|\rho_{1}|^4 |\rho_{2}|^4
\partial_{1} \partial_{2}^{\ast}
\int \frac{d^2 \frho_{0}}
{|\rho_{10}|^2}
\frac{\rho_{0}^{\ast}\rho_{1}}{\rho_{1}^{\ast}\rho_{0}}
\theta(\frac{|\rho_{10}||\rho_{2}| }
{|\rho_{12}||\rho_{0}| } - \epsilon)
\partial_{0}^{\ast} \partial_{2}
\Phi_{2}(\frho_{0},\frho_{2})
\nonumber \\
&&\spc{2cm}+\; 2\pi \ln \epsilon
 |\partial_{1}|^2|\partial_{2}|^2
 \Phi_{2}(\frho_{1},\frho_{2})
\;+\;\left[ \mbox{h.c.} \right]
\;+\; \left[ 1 \leftrightarrow 2 \right].
\eeqn
We rewrite the numerator in front of the $\theta$ - function
as
\beqn
\rho_{0}^{\ast}\rho_{1} & = & (\rho_{01}^{\ast}+
\rho_{1}^{\ast})(\rho_{10}+\rho_{0}) \nonumber \\
 &=& \rho_{01}^{\ast}\rho_{10}+\rho_{01}^{\ast}\rho_{0}
+\rho_{1}^{\ast}\rho_{10} + \rho_{1}^{\ast}\rho_{0}
\eeqn
and consider the four terms seperately.
For the first three terms the singularity in $\frho_0 =\frho_1$ becomes
integrable and the $\theta$ - function can be removed.
We obtain for the first term:
\beqn
|\rho_{1}|^4 |\rho_{2}|^4
\partial_{1} \partial_{2}^{\ast}
\int d^2 \frho_{0}
\frac{-1}{\rho_{1}^{\ast}\rho_{0}}
\partial_{0}^{\ast} \partial_{2}
\Phi_{2}(\frho_{0},\frho_{2})\;
 =\;   |\rho_{1}|^4 |\rho_{2}|^4
 \pi \delta^{(2)} (\frho_{1}) |\partial_2|^2
\Phi_{2}(0,\frho_{2}),
\eeqn
where the rhs emerges after partial integration with respect to $\frho_0$
and the use of the following identities
\beqn
\Delta \ln |\frho|^2\; =\; 4 \partial \partial^* \ln |\rho|^2 \;=\;
                   4 \pi \delta^{(2)}(\frho)
\eeqn
\beqn
\partial \frac{1}{\rho^{\ast}}\;=\;
\partial^{\ast} \frac{1}{\rho}
\;=\;\pi \delta^{(2)} (\frho)\;\;.
\eeqn
The second and third term of (2.28) are calculated in a similar way:
\beqn
|\rho_{1}|^4 |\rho_{2}|^4
\partial_{1} \partial_{2}^{\ast}
\int d^2 \frho_{0}
\left[ \frac{1}{\rho_{1}^{\ast}\rho_{01}} +
\frac{1}{\rho_{0}\rho_{10}^{\ast}} \right]
\partial_{0}^{\ast} \partial_{2}
\Phi_{2}(\frho_{0},\frho_{2}) \nonumber \\
=
|\rho_{1}|^4 |\rho_{2}|^4
\left[ - \partial_1 \frac{1}{\rho_{1}^{\ast}} +
\frac{1}{\rho_{1}}  \partial_{1}^{\ast}  \right]
|\partial_2|^2\Phi_{2}(\frho_{1},\frho_{2})\;\;.
\eeqn
The conjugate expression gives
\beqn
|\rho_{1}|^4 |\rho_{2}|^4
\left[- \partial_1^{\ast}\frac{1}{\rho_{1}} +
\frac{1}{\rho_{1}^{\ast}} \partial_{1} \right]
|\partial_2|^2\Phi_{2}(\frho_{1},\frho_{2}).
\eeqn
After commutation of derivatives we obtain terms
proportional to $\delta^{(2)} (\rho_{1})$ - functions which cancel
against (2.28); the remaining terms in (2.31) and (2.32) add up to zero.
Consequently the only contribution comes from the fourth term of (2.27),
 i.e. after the transformation (2.22) the kernel
(2.21) reads
\beqn
2 \; |\rho_{1}|^4 |\rho_{2}|^4
 |\partial_{1}|^2|\partial_{2}|^2
({\cal K}\Phi)(\frho_{1},\frho_{2}) &=&
|\rho_{1}|^4 |\rho_{2}|^4
\rho_{1}^2 \rho_{2}^{\ast \; 2}
\partial_{1} \partial_{2}^{\ast}
\int \frac{d^2 \frho_{0}}
{|\rho_{10}|^2}
\theta(\frac{|\rho_{10}||\rho_{2}| }
{|\rho_{12}||\rho_{0}| } - \epsilon)
\partial_{0}^{\ast} \partial_{2}
\Phi_{2}(\frho_{0},\frho_{2}) \nonumber \\
&&+\; 2\pi \ln \epsilon
 |\partial_{1}|^2|\partial_{2}|^2
 \Phi_{2}(\frho_{1},\frho_{2})
\;+\;\left[ \mbox{h.c.} \right]
\;+\; \left[ 1 \leftrightarrow 2 \right].
\eeqn
By changing the argument of the $\theta$ - function analogous to
eq.(2.20) the factor $|\rho_2|/|\rho_0|$ in the argument is
eliminated. The extra term $\ln (|\rho_2|/|\rho_1|)$ which arrises
after the change of the argument
cancels due to its antisymmetry under $1 \leftrightarrow 2$.
After removing the overall factor
$|\rho_{1}|^4 |\rho_{2}|^4$, we recover the original expression
(2.21).
\\ \\
\section{The Transition Vertex}
\setcounter{equation}{0}
\noindent
Before we start investigating the symmetry properties of the transition
vertex ~\cite{BW} we present a brief qualitative discussion of how
this vertex is derived; in particular we would like to emphasize that
the derivation of this vertex cannot be understood without
the reggeization of the gluon. \\ \\
Let us first go back to the BFKL kernel. In the previous section
we have already made use of the fact that it is derived from an effective
production vertex (Fig.2a): an s-channel gluon is produced from a
reggeized t-channel gluon, and the ``square`` of this production
vertex leads to the (connected) part of the BFKL kernel.
The iteration of this kernel in the t-channel defines the BFKL Pomeron,
where the t-channel gluons are reggeized. It is then often convenient to
rearrange the sum of ladders by combining the trajectory function of
the t-channel gluons with the BFKL kernel. This gives rise to the
additional (disconnected) piece in the kernel (Fig.2b) which serves as
a regulator in the infrared region, and the t-channel
gluons now come with ordinary propagators $1/k^2$. A similar structure
has been found for the 2 $\rightarrow$ 4 transition vertex. The basic
(connected)
contribution is obtained by taking products of higher order production
vertices (Fig.3a). Disconnected pieces appear if we include other
production mechanisms (an example is given in Fig.3b) and make use of
the fact that, in the three gluon amplitude above the s-channel gluon,
there is a contribution \footnote{As it is shown in ~\cite{B}, the
three gluon amplitude can be written as a sum of three terms,
corresponding to the reggeization of gluon pairs (12), (23), (13),
respectively.} where the two gluon lines on the right reggeize
into a single gluon (Fig.3c). By applying this type of arguments
to all possible production vertices and invoking reggeization for the
three gluon amplitude above, one arives at a 2 $\rightarrow$ 4
transition vertex which consists of connected and disconnected pieces,
in close analogy with the BFKL kernel.\\ \\
However, this is not yet the final form of the vertex. The four gluon
amplitude obtained so far still includes the possibility that
the gluons at the bottom may reggeize, e.g. the two gluons on the
left and the two gluons on the right may form a single gluon each.
It turns out that contributions of this kind have to be {\it subtracted}
from the previous four gluon amplitude, and only the remainder defines
the final 2 $\rightarrow$ 4 transition vertex. \\ \\
Let us then state the form of the transition vertex which we are going
to use. Following ~\cite{BW} we write
\beqn
V^{a_1,a_2,a_3,a_4}(k_1,k_2,k_3,k_4) =
      \delta^{a_1,a_2} \delta^{a_3,a_4} V(k_1,k_2,k_3,k_4)
\nonumber \\
   +  \delta^{a_1,a_3} \delta^{a_2,a_4} V(k_1,k_3,k_2,k_4)
   +  \delta^{a_1,a_4} \delta^{a_2,a_3} V(k_1,k_4,k_2,k_3),
\eeqn
where the $\delta$-symbols refer to the color structure, and the
momentum space expressions for the function
$V(k_1,k_2,k_3,k_4)$ can be found in ~\cite{BW} and will not be
presented here. Instead we make use of the graphical representation
shown in Fig.4 and adopt the following rules:\\
(1) each line with (transverse) momentum $k$ has a propagator
$1/k^2$, each closed loop the integral $\int \frac{d^2k}{(2\pi)^3}$.\\
(2) each vertex carries a factor $q^2$ where $q$ denotes the sum of
momenta above or below the vertex.\\
(3) Sum over the permutations of external momenta, as indicated in
Fig.4.\\
(4) at the end, multiply with the overall factor $\frac{g^4}{4\sqrt{2}}$.
\\
In the following section we shall show that this function $V$ is
invariant under (conformal) M\"obius transformations.\\  \\
It should be remarked that the diagrams in Fig.4
may also be grouped in a different way allowing the use of the
complex notation for the effective emission vertex (or Lipatov vertex).
In particular, when addressing the question of holomorphic separability
the complex notation is needed and
hopefully leads to an operator like approach similar
to that found for the BFKL-kernel in eqs.(1.1)-(1.4).
For the purpose of this paper, the proof of the conformal invariance of $V$,
the representation in Fig.4 turned out to be more convenient.
questions concerning the holomorphic separability will be discussed
elsewhere.

\section{Proof of the conformal invariance}
\setcounter{equation}{0}
Let us first define the expression which we are going to investigate
(Fig.5): the upper blob denotes the two-gluon amplitude $\Phi_2$
of the previous section which includes the $1/k^2$ propagators of the
upper two
internal gluon lines. Correspondingly, we absorb the
propagators of the four lower internal lines into the lower blob which
we denote by $\Phi_4 (k_1, k_2,k_3,k_4)$ (we again suppress the
$\omega$-dependence). Fourier transforms are
defined as in (2.7). We shall make use of the fact ~\cite{L86,L93} that
$\Phi_2$ and $\Phi_4$ have already been shown to be invariant under
conformal
transformations, and we prove that the additional integral which
involves the new transition kernel preserves this symmetry,
in particular the invariance with respect to the inversion. \\ \\
We divide the set of graphs shown in Fig.4 into
the four groups A, B, C, D. The last group D is easily recognized as
the BFKL kernel, with the lower momenta being $k_1+k_2$ and $k_3+k_4$.
The conformal invariance of this expression has been proven in
the previous section. For the other three groups we will proceed
in the same fashion as in section 2: starting from the momentum
space expressions of Fig.4 we first derive the representations in
configuration space
 and then prove the invariance under inversion.
We shall find that none of the subsets A, B, C is invariant by itself;
only the the sum of all terms has the symmetry property.\\ \\
Let us begin with the group C and consider the first line with the
momenta (1234). It consists of three pieces which
in configuration space take the form (the final expression
for $V$ in configuration space has to be completed by an
overall factor $\frac{g^4}{16 \sqrt{2} \pi^2}$):
\beqn
&&\int d^2 \frho_1 d^2 \frho_2 \Phi_4 (\frho_1, \frho_1, \frho_1, \frho_2)
     \frac{1}{2} \delta^{(2)}(\frho_{12}) (\nabla_1 + \nabla_2)^2
     \Phi_2 (\frho_1, \frho_2)\nonumber \\
&+&  \frac{1}{2\pi} \int d^2 \frho_1 d^2 \frho_2 \nabla_1^2
                    \Phi_4(\frho_1, \frho_1, \frho_1,\frho_2)
     \left[\ln \frac{2}{|\rho_{12}|} - \ln \lambda +\psi(1) \right]
                                    \nabla_2^2 \Phi_2 (\frho_1, \frho_2)
 \\
&+&  \int d^2 \frho_1 d^2 \frho_2 d^2 \frho_0
   \Phi_4 (\frho_1, \frho_1, \frho_1, \frho_2) \frac{1}{(2 \pi)^2}
   \left( \frac{\theta (|\rho_{10}| - \epsilon)}{| \rho_{10}|^2}
        +c(\epsilon, \lambda) \delta^{(2)}(\frho_{10})  \right)
               \nabla_0 ^2 \nabla_2^2  \Phi_2 (\frho_0, \frho_2)
\nonumber\eeqn
with  $c(\epsilon, \lambda)$ from (13). One sees that the constant
pieces $\ln 2 - \ln \lambda + \psi(1)$ cancel, and we are left with:
\beqn
&&\int d^2 \frho_1 d^2 \frho_2 \Phi_4 (\frho_1, \frho_1, \frho_1, \frho_2)
     \frac{1}{2} \delta^{(2)}(\frho_{12}) (\nabla_1 + \nabla_2)^2
     \Phi_2 (\frho_1, \frho_2)
\nonumber \\
&-&  \frac{1}{2\pi} \int d^2 \frho_1 d^2 \frho_2 \nabla_1^2
                    \Phi_4(\frho_1, \frho_1, \frho_1,\frho_2)
     \ln |\rho_{12}| \nabla_2^2 \Phi_2 (\frho_1, \frho_2)
 \\
   &+& \frac{1}{(2 \pi)^2}  \int d^2 \frho_1 d^2 \frho_2 d^2 \frho_0
   \Phi_4 (\frho_1, \frho_1, \frho_1, \frho_2)
   \left( \frac{\theta (|\rho_{10}| - \epsilon)}{| \rho_{10}|^2}
        + 2\pi \delta^{(2)} (\frho_{10}) \ln \epsilon \right)
               \nabla_0 ^2 \nabla_2^2  \Phi_2 (\frho_0, \frho_2)\;\;.
 \nonumber
\eeqn
Before we apply the transformation of inversion, it will be convenient
to perform a few changes. First we change the second term. After partial
integration we get (cf.eq.(2.28)):
\beqn
\nabla_1 ^2 \ln |\rho_{12} | \Phi_2 (\frho_{1}, \frho_{2} )
&=& \nabla_1
  \left( \frac { \frho_{12} } {|\rho_{12} |^2 } \Phi_2 (\frho_{1}, \frho_2)
+\ln |\rho_{12}| \nabla_1 \Phi_2 (\frho_1, \frho_2) \right) \\
&=& 2 \pi \delta^{(2)}(\frho_{12}) \Phi_2 (\frho_1, \frho_2)
      + 2 \frac { \frho_{12} } {|\rho_{12} |^2 } \nabla_1
            \Phi_2(\frho_{1},\frho_2)
      +\ln |\rho_{12} | \nabla_1^2 \Phi_2 (\frho_1, \frho_2)\;\;.\nonumber
\eeqn
Here the first term can be combined with the first term in (4.2).
In the third term of (4.2) we split the $\nabla_0^2$ operator:
by partial integration we obtain:
\beqn
\frac{1}{(2 \pi)^2} \int d^2\frho_1 d^2 \frho_2 d^2 \frho_0
\Phi_4 (\frho_1, \frho_1, \frho_1, \frho_2)  \left(
\nabla_1 \frac{ \theta(|\rho_{10}| - \epsilon) }{|\rho_{10}|^2} \nabla_0
    + 2 \pi \delta^{(2)} (\frho_{10}) \ln \epsilon \nabla_0^2 \right)
    \nabla_2^2 \Phi_2(\frho_0, \frho_2)  .
\eeqn
We rescale the argument of the $\theta$-function and obtain the
additional logarithm $2\pi \ln |\rho_{12}|$. After differentiation
we arrive at:
\beqn
&&\frac{1}{(2 \pi)^2} \int d^2\frho_1 d^2 \frho_2 d^2 \frho_0
\Phi_4 (\frho_1, \frho_1, \frho_1, \frho_2) \left(
\nabla_1 \frac{ \theta( \frac{|\rho_{10}|}{|\rho_{12}|} - \epsilon)}
            {|\rho_{10}|^2} \nabla_0
    + 2 \pi \delta^{(2)} (\frho_{10}) \ln \epsilon \nabla_0^2 \right)
    \nabla_2^2 \Phi_2(\frho_0, \frho_2) \nonumber
       \\
    &+& \frac{1}{2 \pi} \int d^2\frho_1 d^2\frho_2 d^2\frho_0
       \Phi_4 (\frho_1, \frho_1, \frho_1, \frho_2) \left(
      \frac{\frho_{12} \nabla_1}{|\rho_{12}|^2}
    +  \ln |\rho_{12}| \nabla_1^2 \right)
    \nabla_2^2 \Phi_2(\frho_1, \frho_2) \;\;.
\eeqn
The next-to-last term combines with a similar term in (4.3), the
last term cancels against the third piece of (4.3). Eq.(4.2) therefore
takes the form:
\beqn
&&\int d^2 \frho_1 d^2 \frho_2 \Phi_4 (\frho_1, \frho_1, \frho_1, \frho_2)
   \left(\delta^{(2)}(\frho_{12}) \nabla_1 \nabla_2
                                 - \frac{1}{2\pi} \frac{\frho_{12}}
    {|\rho_{12}|^2} \nabla_1  \nabla_2^2 \right) \Phi_2 f(\frho_1,\frho_2)
   \nonumber  \\
&+& \frac{1}{(2 \pi)^2} \int d^2 \frho_1 d^2 \frho_2 d^2 \frho_0
           \Phi_4 (\frho_1, \frho_1, \frho_1,\frho_2)\;\;\cdot\\
&&\spc{3cm}\cdot\;\;
 \left( \nabla_1 \frac{\theta(\frac{|\rho_{10}|}{|\rho_{12}|} - \epsilon)}
               { |\rho_{10}|^2} \nabla_0
                + 2\pi \delta^{(2)}(\frho_{10}) \ln \epsilon
      \nabla_0^2 \right)
                 \nabla_2^2 \Phi_2 (\frho_0, \frho_2)\;\;.\nonumber
\eeqn
\\ \\
As in the case of the BFKL kernel it will now be convenient to switch
to the complex notation. With $\nabla^2 = 4 \partial
\partial^*$ eq.(4.6) becomes:
\beqn
&&4\; \int d^2 \frho_1 d^2 \frho_2 \Phi_4 (\frho_1, \frho_1, \frho_1, \frho_2)
   \left(\frac{1}{2} \delta^{(2)}(\frho_{12}) (\partial_1 \partial_2^*
        +\partial_1^* \partial_2)
 - \frac{1}{2\pi}(\frac{1}{\rho_{12}} \partial_1^* + h.c.)
        |\partial_2|^2 \right)  \Phi_2 (\rho_1,\rho_2)
             \nonumber \\
&+& \frac{2}{\pi^2} \int d^2 \frho_1 d^2 \frho_2 d^2 \frho_0
           \Phi_4 (\frho_1, \frho_1, \frho_1,\frho_2) \;\cdot\\
&&\spc{3cm}\cdot\;\left( \partial_1
  \frac{\theta(\frac{|\rho_{10}|}{|\rho_{12}|}- \epsilon)}
               {  |\rho_{10}|^2} \partial_0^*\; +\; \mbox{h.c.}
\; + \;4\pi \delta^{(2)}(\frho_{10}) \ln \epsilon
          |\partial_0|^2 \right)
                     \;       |\partial_2|^2 \Phi_2 (\frho_0, \frho_2)\;\;.
\nonumber\eeqn
This represents the configuration space expression for the first term
of the group C. \\ \\
Under the inversion transformation $\rho \to 1/\rho$,
$\rho^* \to 1/ \rho^*$ with
\beqn
\delta^{(2)}(\frho_{12}) \to |\rho_1|^2 |\rho_2|^2
      \delta^{(2)}(\frho_{12})
\eeqn
eq.(4.3) turns into:
\beqn
&4& \int d^2 \frho_1 d^2 \frho_2 \Phi_4 (\frho_1, \frho_1, \frho_1, \frho_2)
   \left(\frac{1}{2} \delta^{(2)}(\frho_{12}) (\partial_1 \partial_2^*
       +\partial_1^* \partial_2)  - \frac{1}{2\pi}(\frac{\rho_2}
   {\rho_{12} \rho_1} \partial_1^* \;+\; \mbox{h.c.})
        |\partial_2|^2 \right) \Phi_2 (\frho_1,\frho_2)
             \nonumber \\
&+& \frac{2}{\pi^2} \int d^2 \frho_1 d^2 \frho_2 d^2 \frho_0
           \Phi_4 (\frho_1, \frho_1, \frho_1,\frho_2)\;\cdot\\
&&\spc{2cm}\cdot\; \left(\partial_1
        \frac{\theta(\frac{|\rho_{10}| |\rho_1| |\rho_2|}
                          {|\rho_{12}| |\rho_0|^2 } - \epsilon)}
               { |\rho_{10}|^2}
             \frac{\rho_0^* \rho_1}{\rho_0 \rho_1^*} \partial_0^*
       \;+\; \mbox{h.c.}
\;+\; 4\pi \delta^{(2)}(\frho_{10}) \ln \epsilon\;
     |\partial_0|^2  \right)
\; |\partial_2|^2 \Phi_2 (\frho_0, \frho_2)\;\;. \nonumber
\eeqn
In the second line we write:
\beqn
             \frac{\rho_0^* \rho_1}{|\rho_{10}|^2 \rho_0 \rho_1^*} & = &
\frac{(\rho_{10} + \rho_0) (\rho_1 - \rho_{10})^*}
        {|\rho_{10}|^2 \rho_1^* \rho_{0}} \nonumber \\
&=&-\frac{1}{\rho_1^* \rho_0} + \frac{1}{\rho_{10}^* \rho_0} -
\frac{1}{\rho_{10} \rho_1^*} + \frac{1}{|\rho_{10}|^2}.
\eeqn
In the first three terms the $\theta$-function can be ignored
since the logarithmic divergence has disappeared. In the first term
we make use of (2.29) and (2.30) and do the $\rho_0$ integral via
partial integration:
\beqn
\int d^2 \frho_0 [- \partial_1 \frac{1}{\rho_1^*\rho_0} \partial_0^*
+ \mbox{h.c.}] |\partial_2|^2 \Phi_2 (\frho_0, \frho_2)
= 2 \pi^2 \delta^{(2)} (\frho_1) |\partial_2|^2 \Phi_2 (0,\frho_2).
\eeqn
Similarly the  second and the third term lead to:
\beqn
\pi [ \frac{1} {\rho_1} \partial_1^* + \mbox{h.c.}] |\partial_2|^2 \Phi_2
        (\frho_1, \frho_2)
\eeqn
and
\beqn
-2 \pi^2 \delta^{(2)} (\frho_1) |\partial_2|^2 \Phi_2(\rho_1,\rho_2)
- \pi [\frac{1}{\rho_1^*} \partial_1^* + \mbox{h.c.}] |\partial_2|^2 \Phi_2
         (\frho_1, \frho_2) ,
\eeqn
resp.. One easily sees that the first piece in (4.13) cancels against
(4.11), the second piece in (4.13) against (4.12). So all we are left
with is the fourth term in (4.10). Here we again rescale the argument of
the $\theta$ function, in order to compare with our starting expression
(4.7). The additional logarithmic term has the form:
\beqn
&&\frac{4}{\pi} \int d^2\frho_1 d^2\frho_2  d^2\frho_0
       \Phi_4(\frho_1,\frho_1,\frho_1, \frho_2) \left[ \partial_1 \ln
        \frac{|\rho_2|}{|\rho_1|} \partial_1^* + \mbox{h.c.}
\right] |\partial_2|^2
             \Phi_2(\frho_1, \frho_2)
 \\ & =&
\frac{4}{\pi} \int d^2\frho_1 d^2\frho_2 d^2\frho_0
       \Phi_4(\frho_1,\frho_1,\frho_1, \frho_2) \left[
      -\frac{1}{\rho_1} \partial_1^* - \mbox{h.c.}
      + 2 \ln \frac{|\rho_2|}{|\rho_1|}
                             |\partial_1|^2 \right] |\partial_2|^2
             \Phi_2(\frho_1, \frho_2)\;\;.\nonumber
\eeqn
Finally, in the first line of (4.9) we put
\beqn
\frac{\rho_2}{\rho_{12} \rho_1} = \frac{1}{\rho_{12}} -
                        \frac{1}{\rho_1}.
\eeqn
and observe that the second term cancels against the first part of
the second line of (4.14). As a result of the inversion, we arrive at
our old result plus the extra term:
\beqn
\frac{8}{\pi} \int d^2\frho_1 d^2\frho_2 \Phi(\frho_1, \frho_1, \frho_1,
         \rho_2) \ln \frac{|\rho_2|}{|\rho_1|} |\partial_1|^2
         |\partial_2|^2 \Phi_2 (\frho_1, \frho_2).
\eeqn
Later on we shall see that this term will be cancelled by similar terms
from A and B.
\\ \\
Next we turn to group A and consider the momentum set (1234). In
configuration space we have:
\beqn
&&\spc{1cm} \int d^2\frho_1 d^2\frho_2 d^2\frho_0
\Phi_4 (\frho_1, \frho_0, \frho_0, \frho_2) \nabla_0^2
\frac{\ln |\rho_{10}|  \ln |\rho_{20}| }{(2 \pi)^2}
\nabla_1^2 \nabla_2^2 \Phi_2 (\rho_1, \rho_2)   \\
    &-&\frac{1}{(2\pi)^2} \int d^2 \frho_1 d^2 \frho_2 d^2\frho_0
\Phi_4 (\frho_1, \frho_1, \frho_1, \frho_2)
      \left( \frac{ \theta(|\rho_{10}| - \epsilon)}{|\rho_{10}|^2}
         +2\pi \delta^{(2)} (\frho_{10}) \ln \epsilon \right)
      \; \nabla_1^2 \nabla_2^2 \Phi_2 (\frho_1, \frho_2)
       \nonumber  \\
     &-&\frac{1}{(2\pi)^2} \int d^2 \frho_1 d^2 \frho_2 d^2\frho_0
\Phi_4 (\frho_1, \frho_2, \frho_2, \frho_2)
      \left( \frac{ \theta(|\rho_{20}| - \epsilon)}{|\rho_{20}|^2}
         +2\pi \delta^{(2)} (\frho_{20}) \ln \epsilon \right)
     \;  \nabla_1^2 \nabla_2^2 \Phi_2 (\frho_1, \frho_2)\;\;.\nonumber
\eeqn
Here we have already omitted the constant pieces $\ln2 - \ln \lambda
+\psi(1)$: they appear both in the first line (in combination with
$\ln |\rho_{10}| $) and the trajectory function term in the second
and third term, and in the sum they cancel.
In the first term we take the derivative with respect to $\rho_0$:
\beqn
\nabla_0^2 \ln |\rho_{10}| \ln |\rho_{20}|
= 2 \pi \delta^{(2)} (\frho_{10}) \ln |\rho_{20}|
+ 2 \pi \delta^{(2)} (\frho_{20}) \ln |\rho_{10}|
    + 2 \frac{\frho_{10} \cdot \frho_{20} }
           {|\rho_{10}|^2 |\rho_{20}|^2} .
\eeqn
In the first two terms we use the $\delta$-functions and perform the
$\rho_0$ integral. The third term can be combined with the $\theta$-
function pieces in (4.17) which we rewrite in a symmetric form
$\theta(|\rho_{10}| - \epsilon)= \theta(\frac{|\rho_{10}||\rho_{20}|}
{|\rho_{12}|} -\epsilon)$. The result is:
\beqn
&&\frac{1}{(2\pi)} \int d^2\frho_1 d^2 \frho_2 \;\left(
     \Phi_4 (\frho_1,\frho_1,\frho_1,\frho_2) \ln |\rho_{12}| +
     \Phi_4 (\frho_1,\frho_2,\frho_2,\frho_2) \ln |\rho_{12}| \right)
 \nonumber \\
&-& \frac{1}{(2\pi)^2} \int d^2\frho_1 d^2\frho_2 d^2\frho_0\;\;
       \Phi_4 (\frho_1,\frho_0,\frho_0,\frho_2) \;\cdot\\
&&\spc{1cm}\cdot\;  \left(
    \frac{|\rho_{12}|^2}{|\rho_{10}|^2 |\rho_{20}|^2}
    \theta(\frac{|\rho_{10}||\rho_{20}|}{|\rho_{12}|} - \epsilon)\;
 +\; 2\pi \ln \epsilon \left[\delta^{(2)}(\frho_{10})
+ \delta^{(2)}(\frho_{20}) \right]
\right) \nabla_1^2 \nabla_2^2 \Phi_2 (\frho_1, \frho_2) \;\;.\nonumber
\eeqn
The first line can be absorbed into the $\theta$-functions (cf.
(2.21)). So we are left with:
\beqn
&-& \frac{1}{(2\pi)^2} \int d^2\frho_1 d^2\frho_2 d^2\frho_0
       \Phi_4 (\frho_1,\frho_0,\frho_0,\frho_2) \;\cdot \\
&&\spc{1cm}  \cdot\; \left(
    \frac{|\rho_{12}|^2}{|\rho_{10}|^2 |\rho_{20}|^2}
    \theta(\frac{|\rho_{10}||\rho_{20}|}{|\rho_{12}|^2} - \epsilon)\;
+\;2\pi \ln \epsilon \left[\delta^{(2)}(\frho_{10}) + \delta^{(2)}(\frho_{20})
\right]\right) \nabla_1^2 \nabla_2^2 \Phi_2 (\frho_1, \frho_2) \;\;.\nonumber
\eeqn
We again switch to the complex notation:
\beqn
&-& \frac{4}{\pi^2} \int d^2\frho_1 d^2\frho_2 d^2\frho_0
       \Phi_4 (\frho_1,\frho_0,\frho_0,\frho_2)\;\cdot\\
&&\spc{1cm}\cdot\;  \left(
    \frac{|\rho_{12}|^2}{|\rho_{10}|^2 |\rho_{20}|^2}
    \theta(\frac{|\rho_{10}||\rho_{20}|}{|\rho_{12}|^2} - \epsilon)
+ 2\pi \ln \epsilon \left[\delta^{(2)}(\frho_{10}) + \delta^{(2)}(\frho_{20})
\right]\right)
        |\partial_1|^2 |\partial_2|^2 \Phi_2 (\frho_1, \frho_2)\;\;.\nonumber
\eeqn
This is the configuration space expression of the first term in A.
\\ \\
Under the inversion only the $\theta$-function requires a little
calculation. Its argument behaves as:
\beqn
\frac{|\rho_{10}| |\rho_{20}|}{|\rho_{12}|^2} \to
\frac{|\rho_{10}| |\rho_{20}|}{|\rho_{12}|^2}
        \frac{|\rho_1| |\rho_2|}{|\rho_0|^2} ,
\eeqn
which is equivalent to having the extra term
\beqn
-\frac{8}{\pi} \int d^2\frho_1 d^2 \frho_2 \left(
     \Phi_4 (\frho_1,\frho_1,\frho_1,\frho_2)
                  \ln \frac{|\rho_{2}|}{|\rho_1|}
   + \Phi_4 (\frho_1,\frho_2,\frho_2,\frho_2)
                 \ln \frac{|\rho_{1}|}{|\rho_2|} \right)
|\partial_1|^2 |\partial_2|^2 \Phi_2 (\frho_1,\frho_2).
\eeqn
\\ \\
Finally we come to group B (first line in Fig.4, momenta (1234)). It
differs from A by the overall sign and the arguments of $\Phi_4$
which are replaced by $\Phi_4 (\frho_1, \frho_1, \frho_0, \frho_2)$.
Repeating the same sequence of steps we find that, under inversion,
the additional terms analogous to (4.23) are:
\beqn
\frac{8}{\pi} \int d^2\frho_1 d^2 \frho_2 \left(
     \Phi_4 (\frho_1,\frho_1,\frho_1,\frho_2)
                  \ln \frac{|\rho_{2}|}{|\rho_1|}+
     \Phi_4 (\frho_1,\frho_1,\frho_2,\frho_2)
                 \ln \frac{|\rho_{1}|}{|\rho_2|} \right)
|\partial_1|^2 |\partial_2|^2 \Phi_2 (\frho_1,\frho_2)  .
\eeqn
\\ \\
One notices that the remainders of A, B, and C
(in (4.23), (4.24), and (4.16),
respectively) are all of the same form but differ in the arguments
of $\Phi_4$. As a final step, we have to include the permutations of
the arguments, as indicated in Fig.4. In an obvious notation, we have:
\beqn
A: & [1234] + [2134] + [1243] + [2143]  \nonumber \\
B: & [1234] + [1243] + second \,\,line
                     \nonumber \\
C: & [1234] + [1243] + second \,\,line .
\eeqn
Inserting these changes of arguments into (4.23) we obtain for A:
\beqn
&-&\frac{16}{\pi} \int d^2\frho_1 d^2\frho_2 \left(
[\Phi_4 (\frho_1,\frho_1,\frho_1,\frho_2) +
     \Phi_4 (\frho_1,\frho_1,\frho_2,\frho_1)]
\ln \frac{|\rho_2|}{|\rho_1|}  \right.   \\
&&\spc{2cm}\left. +\;
[\Phi_4 (\frho_1,\frho_2,\frho_2,\frho_2) +
     \Phi_4 (\frho_2,\frho_1,\frho_2,\frho_2)]
\ln \frac{|\rho_1|}{|\rho_2|}  \right)
|\partial_1|^2 |\partial_2|^2 \Phi_2 (\frho_1,\frho_2) \;\;.\nonumber
\eeqn
Similarly for B in (4.24):
\beqn
&&\frac{8}{\pi} \int d^2\frho_1 d^2\frho_2 \left(
[\Phi_4 (\frho_1,\frho_1,\frho_1,\frho_2) +
     \Phi_4 (\frho_1,\frho_1,\frho_2,\frho_1)
        +2\Phi_4 (\frho_1,\frho_1,\frho_2,\frho_2)]
\ln \frac{|\rho_2|}{|\rho_1|}  \right. \\
&&\left. +\;
[2\Phi_4 (\frho_1,\frho_1,\frho_2,\frho_2)+
 \Phi_4 (\frho_1,\frho_2,\frho_2,\frho_2) +
     \Phi_4 (\frho_2,\frho_1,\frho_2,\frho_2)]
\ln \frac{|\rho_1|}{|\rho_2|} \right)
|\partial_1|^2 |\partial_2|^2 \Phi_2 (\frho_1,\frho_2)\;\; .\nonumber
\eeqn
In the same way we find for the remainders of C in (4.16):
\beqn
&&\frac{8}{\pi} \int d^2\frho_1 d^2\frho_2 \left(
[\Phi_4 (\frho_1,\frho_1,\frho_1,\frho_2) +
     \Phi_4 (\frho_1,\frho_1,\frho_2,\frho_1)]
\ln \frac{|\rho_2|}{|\rho_1|}  \right. \\
&&\spc{2cm} \left. +\;
[\Phi_4 (\frho_1,\frho_2,\frho_2,\frho_2) +
     \Phi_4 (\frho_2,\frho_1,\frho_2,\frho_2)]
\ln \frac{|\rho_1|}{|\rho_2|}  \right)
|\partial_1|^2 |\partial_2|^2 \Phi_2 (\frho_1,\frho_2)\;\; .\nonumber
\eeqn
Taking the sum of (4.26) and (4.27) and (4.28), we arrive at zero. This
completes our proof that the sum of the terms A, B, C is
invariant under the inversion of coordinates. As it was said before,
the last group D has the form of the BFKL kernel and is therefore
invariant by itself.

\section{Conclusions}
\setcounter{equation}{0}
The result of this paper is the invariance of the effective
2 $\to$ 4 gluon vertex under M\"obius transformations, in particular
the inversion $\rho \to 1/\rho$. As a consequence, one expects that
the quantum mechanical problem of n pairwise-interacting gluons
which due to the existence of the number nonconserving vertex now turns
into a quantum field theory preserves the conformal invariance.
As the next step it is extremely important to find out whether the
transition vertex is separable, i.e. it can be written as a sum
of two terms which depend only upon either the $\rho$'s or the
$\rho^*$'s. It is, therefore, very desirable to find a compact
form in analogy to (1.1) - (1.4). Work in this direction is in progress.
\\ \\

{\bf Acknowledgements:} One of us (L.L.) wishes to thank DESY and
the II.Institut f\"{u}r\\ Theoretische Physik, Universit\"{a}t Hamburg,
for their hospitality and the Alexander von\\ Humboldt Foundation
for the financial support. We thank H.Lotter for helpful discussions
on the proof of conformal invariance of the BFKL kernel.

\newpage
\section*{Figure captions}
\noindent
{\bf Fig.1:} The BFKL equation (2.6).\\ \\
{\bf Fig.2:} The structure of the BFKL kernel\\
(a) real production in the BFKL kernel; (b) the effective BFKL
kernel.\\ \\
{\bf Fig.3:} The structure of the 2 $\to$ 4 transition kernel\\
(a) real production which leads to a connected contribution;
(b) real production which gives a disconnected contribution;
(c) reggeization and the 2 $\to$ 3 transition vertex. The dashed line
denotes the s-channel cutting.
(d) structure of the full 2 $\to$ 4 transition vertex (shown are the
contributions from (a) and (c)).\\ \\
{\bf Fig.4:} The complete representation of the
2 $\to$ 4 transition vertex.\\ \\
{\bf Fig.5:} The transition vertex combined with
the non-amputated amplitudes for 2 and 4 gluons, $\Phi_2$ and $\Phi_4$.

\vspace{4cm}
\begin{figure}[htbp]
  \begin{center}
    \leavevmode
     \input{conform1_30.pstex_t}
  \end{center}
  \caption{}
  \label{fig1}
\end{figure}

\begin{figure}[htbp]
  \begin{center}
    \leavevmode
     \input{conform2_30.pstex_t}
  \end{center}
  \caption{}
  \label{fig2}
\end{figure}

\begin{figure}[htbp]
  \begin{center}
    \leavevmode
     \input{conform3_30.pstex_t}
  \end{center}
  \caption{}
  \label{fig3}
\end{figure}

\begin{figure}[htbp]
  \begin{center}
    \leavevmode
     \input{conform4_27.pstex_t}
  \end{center}
  \caption{}
  \label{fig4}
\end{figure}

\begin{figure}[htbp]
  \begin{center}
    \leavevmode
     \input{conform5_40.pstex_t}
  \end{center}
  \caption{}
  \label{fig5}
\end{figure}

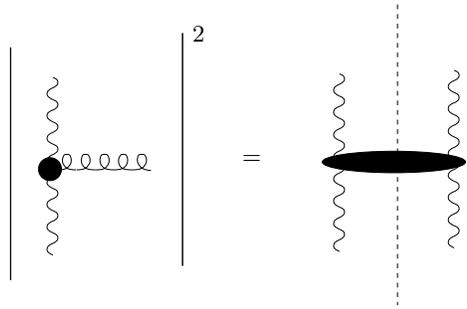
\end{document}